\documentclass[final,3p,times,twocolumn]{elsarticle}

\usepackage{lineno,hyperref}
\usepackage[utf8]{inputenc}
\modulolinenumbers[5]

\journal{Nucl. Instr. and Methods in Phys. Res. Sec. B}

\begin{document}

\begin{frontmatter}

\title{CATLIFE (Complementary Arm for Target LIke FragmEnts): Detector for Target like fragments at VAMOS++ }

\author[SNU,INPA,cens]{Y.~Son}
\author[cens]{Y.~H.~Kim\corref{cor1}}
\cortext[cor1]{Corresponding author}
\ead{yunghee.kim@ibs.re.kr}
\author[SNU,INPA,cens]{Y.~Cho}
\author[SNU,INPA]{S.~Choi}
\author[cens]{J.~Park}
\author[cens]{S.~Bae}
\author[cens]{K.~I.~Hahn}
\author[GANIL]{A.~Navin}
\author[GANIL]{A.~Lemasson}
\author[GANIL]{M.~Rejmund}
\author[GANIL]{D.~Ramos}
\author[GANIL]{E. Cl\'ement} 
\author[GANIL]{D.~Ackermann}
\author[GANIL]{A.~Utepov}
\author[GANIL]{C.~Fougeres}
\author[GANIL]{J. C.~Thomas}
\author[GANIL]{J.~Goupil}
\author[GANIL]{G.~Fremont}
\author[GANIL]{G.~de France}

\address[SNU]{Department of Physics and Astronomy, Seoul National University, Seoul 08826, Republic of Korea}
\address[INPA]{Institute for Nuclear and Particle Astrophysics, Seoul National University, Seoul 08826, Repulic of Korea}
\address[cens]{Center for Exotic Nuclear Studies, Institute for Basic Science, Daejeon 34126, Republic of Korea}
\address[GANIL]{Grand Accélérateur National d'Ions Lourds (GANIL), CEA/DRF-CNRS/IN2P3, F-14076 CAEN Cedex 05, France}

\begin{abstract}
The multi-nucleon transfer reaction between $^{136}$Xe beam and $^{198}$Pt target at the beam energy $7$~MeV/u was studied 
 using the large acceptance spectrometer VAMOS++ coupled with the newly installed second arm time-of-flight and delayed $\gamma$-ray detector CATLIFE (Complementary Arm for Target LIke FragmEnts). 
The CATLIFE detector is composed of a large area multi-wire proportional chamber and the EXOGAM HPGe clover detectors with an ion flight length of 1230 mm.
 Direct measurement of the target-like fragments (TLF) and the delayed $\gamma$-rays from the isomeric state helps to improve TLF identification.
 The use of the velocity of TLFs and the delayed $\gamma$-ray demonstrates the proof of principle and effectiveness of the new setup. 
\end{abstract}


\end{frontmatter}

\section{Introduction}
Multi-nucleon transfer (MNT) reactions between heavy ions are predicted to have an advantage in accessing neutron-rich nuclides near N$=$126 shell closure and super heavy nuclides which are difficult to produce using conventional reaction mechanisms such as fragmentation and fusion evaporation ~\cite{Zagre2008,Dasso1994}. 
Recent experimental results show evidence of the advantage of MNT reaction over conventional fragmentation reaction~\cite{Watanabe2015, Desai2020, Kozulin2012} for the production of nuclei around N$=$126. 
The particle identification of very heavy ions at such low kinetic energies  is very challenging.
Hence only indirect methods such as i) using a statistical model coupled with the kinematics of Projectile-Like Fragments (PLF)~\cite{Watanabe2015} ii) in-beam/decay $\gamma$-ray spectroscopy from of known isotopes~\cite{Desai2020} were used to deduce the cross-sections of the Target-Like Fragments (TLF) in these studies.
The direct kinematic measurement of a TLF can be used to determine the pre-evaporation mass of the TLF~\cite{Kozulin2012,Galtarossa2018} and to understand the corresponding reaction mechanism~\cite{Mija2020}. The presence of isomeric states in the nuclei of interest further helps in their identification. Until now, there is no direct detection of TLF in this region of the nuclear chart. Here we present the first results for CATLIFE (Complementary Arm for Target LIke FragmEnts) 
towards the identification of TLF through the measurement of the velocity and the angular corrections with the PLF and delayed $\gamma$ rays.

\section{Experimental Setup}
The MNT reaction between $^{136}$Xe beam ($7$ MeV/u) and $^{198}$Pt target ($1.3$ mg/cm$^2$) was performed to populate neutron-rich nuclides towards the N$=$126 shell closure to investigate the evolution of nuclear structure in this region. The experiment was carried out at GANIL using the large acceptance spectrometer VAMOS++ ~\cite{REJMUND2011, Kim2017}, positioned at $40^\circ$ with respect to the beam direction, detecting the PLFs with unambiguous particle identification and kinematics measurement.
The prompt $\gamma$-rays from PLFs and TLFs are detected by AGATA (Advanced GAmma Tracking Array), composed of 36 HPGe crystals~\cite{AKKOYUN201226, CLEMENT20171}.   
The array was located at the nominal position (target to the front of the cryostat distance is 230 mm) ~\cite{CLEMENT20171} at backward angles with respect to the beam direction.

\begin{figure}[t]
\includegraphics[width=\columnwidth]{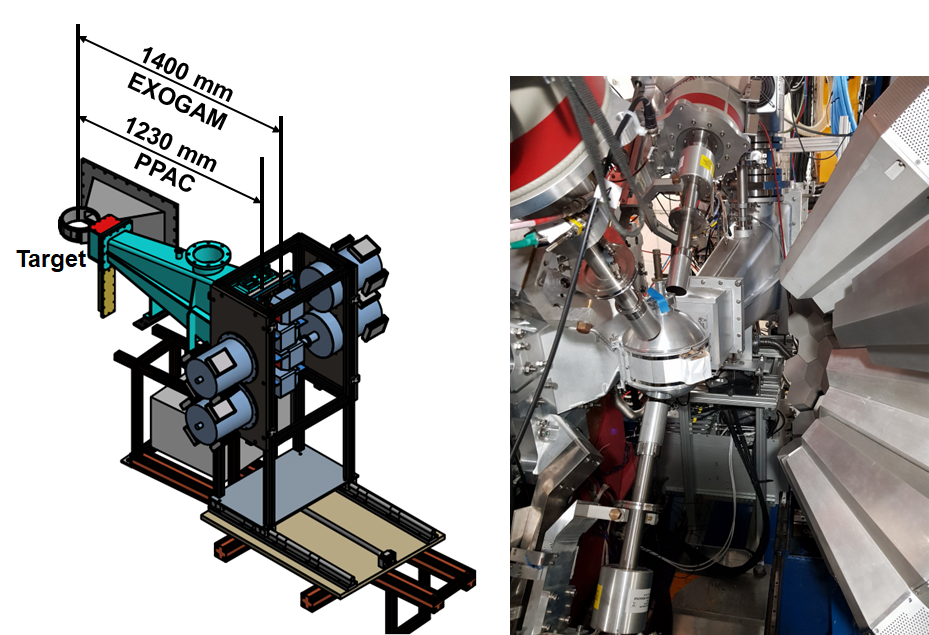}
\caption{(left) The 3D model of the CATLIFE detector with the distance from the target to each component. (right) The picture of the CATLIFE installed at the target position, along with the entrance of the VAMOS++ spectrometer and the AGATA array.} 
\label{fig:setup}
\end{figure}

The CATLIFE detector consists of a target/flight-path vacuum chamber, a multi-wire proportional chamber (MWPC) followed by an Al stopper, and four EXOGAM HPGe clover detectors at the end (see Fig. ~\ref{fig:setup}).

The MWPC is composed of a central cathode wire plane that provides the time signal and two orthogonal anode wires that measure the x and y positions.
The distance between the target to the centre of the MWPC was $1230$ mm.
The area of the MWPC of the CATLIFE with a dimension of $160\times446$ mm$^2$ covering angles between $51^{\circ}$ to $59^{\circ}$ has a geometrical acceptance of $46.5$~msr, which is $58\%$ of the acceptance of VAMOS$++$ ~\cite{REJMUND2011}.
The time-of-flight of the TLFs was measured between the DPS-MWPC (Dual Position Sensitive-MWPC) at the entrance of the VAMOS++~\cite{VAND2016} and the MWPC at the CATLIFE spectrometer. 
The ions exiting the MWPC were implanted in the 2 mm thick Al stopper located behind the exit of the MWPC.

Four EXOGAM Clover HPGe detectors~\cite{simpson2000exogam}, installed $6$ mm behind the 2 mm thick Al stopper at the exit of the MWPC, were used to measure delayed $\gamma$ rays from isomeric states for the  stopped ions (see Fig. ~\ref{fig:setup}). 
The $\gamma$-ray events were recorded using the NUMEXO2 digital electronics with a time window of $0.1~\mu$s to $100~\mu$s from the ion implantation.

\section{Experimental Results}
\subsection{Detection of Target-Like Fragment}

\begin{figure}[h]
\centering
\includegraphics[width=\columnwidth]{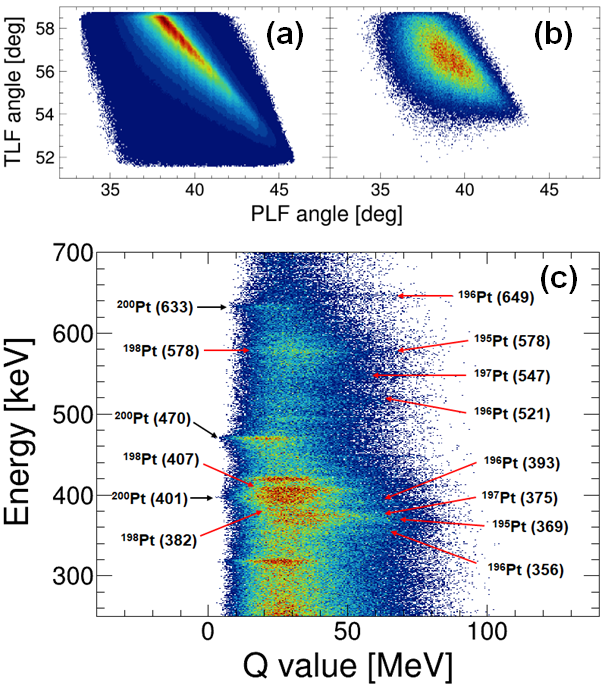}
\caption{Angular correlations between  (a)$^{134}$Xe , (b)$^{138}$Ba PLF in VAMOS++ and TLF in CATLIFE. (c) Doppler corrected prompt $\gamma$-ray energy of TLFs as a function of Q-value corresponding to $^{134}$Xe as PLF. Transitions of different Pt isotopes are indicated.}
\label{fig:AngleCorrel}
\end{figure}

The position resolution of the MWPC was measured to be 300 $\mu$m, using a wire positioned diagonally across the detector. 
The PLF was unambiguously identified in mass (A), charge state (Q) and atomic number (Z)   with a good resolution ($\Delta$Q/Q$\sim$1/86) in VAMOS++ using  machine learning techniques ~\cite{Cho2023}.  The  angular correlation between PLF and TLF for the various  quasi-elastic and deep-inelastic events are shown in Fig.~\ref{fig:AngleCorrel}, with $^{134}$Xe as PLF in (a) and $^{138}$Ba as PLF in (b).

The Q-value for a given transfer reaction was  obtained from the measured  velocities of both PLF (VAMOS++) and TLF (CATLIFE) taking into account  the energy loss correction in the target and assuming that the reaction occurred at the center of the target.
Fig.~\ref{fig:AngleCorrel}(c) shows  the prompt $\gamma$-ray energy gated by the  $^{134}$Xe Doppler corrected for the  TLF partner detected in CATLIFE. A clear effect of neutron evaporation as a function of the Q value can be seen. The corresponding TLF $\gamma$ transitions corresponding to the 0n to 3n evaporation channel as the excitation energy increases can be observed.
The measured Q-value spectrum can be used to select the TLF events with or without neutron evaporation. The deduction of the A  distribution of the TLF before evaporation using the measured velocity is in progress.

\subsection{Delayed $\gamma$-ray spectroscopy}
The delayed $\gamma$ rays from the decay of isomeric states of the TLF implanted in the Al stopper located at the end of CATLIFE are detected by the EXOGAM detectors. To compensate for the effect of Compton scattering, signals from crystals in the same clover were added back. To reduce the background, only the first column  of crystal segments close to the Al catcher was used for added back energy.

\begin{figure}[h]
\includegraphics[width=\columnwidth]{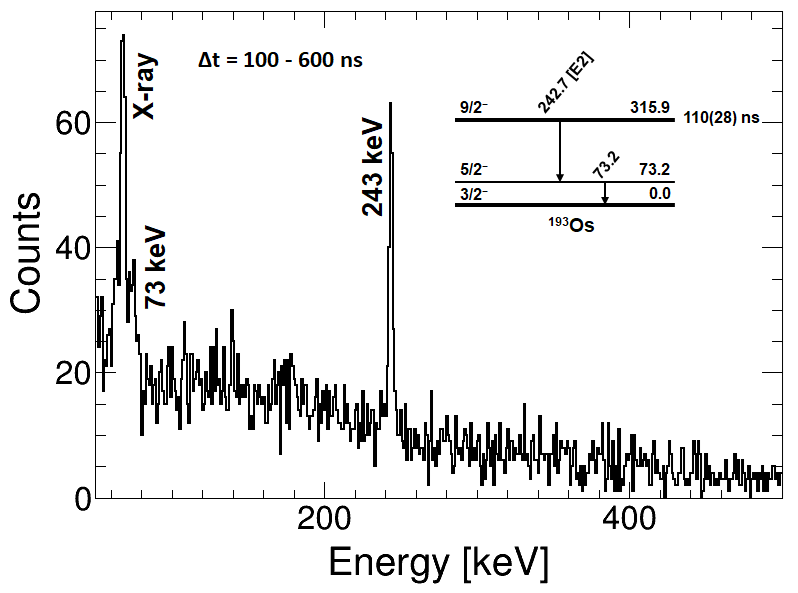}
\caption{Delayed $\gamma$-ray spectrum of $^{193}$Os obtained from  $^{138}$Ba identified in VAMOS++  for  $100\sim600$ ns time window. The $243$ keV and $73.2 $ transitions can be clearly seen.}
\label{fig:Os193}
\end{figure}

The delayed $\gamma$-ray transitions in  TLF from isomeric states could be obtained by gating on a PLF identified by VAMOS++ using a suitable time window related to the half-life of the isomer. 
Fig. ~\ref{fig:Os193}, for example, shows that transitions arising from the isomeric decay of the $9/2^-$ state in $^{193}$Os can be clearly observed by selecting the 3n evaporation channel $^{138}$Ba and a $100$ ns to $600$ ns time window condition. In the same way, using other PLF gates and time window conditions, spectra of the corresponding TLFs can be observed. Clearer spectra can be obtained through background reduction, which is currently in progress.

\section{Summary}

The measurement A-, Z-gated prompt $\gamma$-rays using VAMOS++ and AGATA for PLFs along with the velocity and delayed $\gamma$-rays for the corresponding TLFs formed in MNT reactions in  $ ^{136}$Xe +  $^{198}$Pt system  at 7 MeV/u, using the MWPC and four EXOGAM clovers, is reported. This result demonstrates the  successful implementation of the new CATLIFE spectrometer for  the identification of TLFs  using  the time-of-flight and delayed $\gamma$-ray spectrometer. The Q-value, determined from velocity measurements in VAMOS++ and CATLIFE, in a representative 2n transfer channel ($^{134}$Xe), illustrated  the effect of neutron evaporation in the TLF, using  an  A-, Z-gated PLF prompt $\gamma$-ray spectrum. The measurement of isomeric transitions in $^{193}$Os in two proton pick-up reactions having T$_{1/2}\sim 100$ ns by measuring delayed $\gamma$-rays using EXOGAM HPGe clover detectors were also reported.
Ongoing  analysis for A distribution before neutron evaporation and a further reduction of delayed $\gamma$-ray background will further improve the TLF identification and the corresponding delayed $\gamma$-ray spectroscopy.

\section*{Acknowledgement}
The authors thank for the (financial) support of the National Research Foundation of Korea, South Korea (NRF) grants funded by the Korean government (MSIT), Grants No. 2022R1A2C2005093, No. 2019K2A9A2A10017576, No. 2020R1A2C1005981;
the Institute for Basic Science, South Korea (IBS), Grants No. IBS-R031-D1;
European Union’s Horizon 2020 research and innovation programme under grant agreement No. 654002. We express our gratitude for the support of the GANIL staff and the AGATA collaboration.

\bibliography{cas-refs}

\end{document}